\documentclass[12pt,onecolumn]{article}%
\usepackage{amssymb}
\usepackage{amsmath}
\usepackage{amsfonts}
\usepackage{sw20ssur}
\usepackage{graphicx}%
\newtheorem{theorem}{Theorem}

\newtheorem{lemma}[theorem]{Lemma}

\newtheorem{proposition}[theorem]{Proposition}

\begin{document}

\title{A black hole cast on a non-commutative background}
\author{Manasse R. Mbonye\\\ \textit{Department of Physics, }\\\textit{Rochester Institute of Technology, }\\\textit{85 Lomb Drive, Rochester, NY 14623.} }
\maketitle

\begin{abstract}
In this work we describe a black hole, set on a non-commutative background.
The model, which is relatively simple, is an exact solution of the Einstein
Field Equations. Based on a proposition we put forward, we argue that
introducing a matter density field\ on a non-commutative background sets up a
mechanism that deforms the field into two distinct fields, one residing
dominantly on the lattice tops (hereafter, on-cell) and the other residing
dominantly in the inter-lattice regions (hereafter, off-cell). The two fields
have different physical and themodynamic characterics which we describe, and
some of which play a role in halting collpse to a singularity. For example,
not surprisingly the on-cell (off-cell) fields manifest standard on-shell
(off-shell) characteristics, respectively. Both the density and the net
mass-energy are unchanged by the deformation mechanism. In our treatment the
mass of a black hole defines its own size scale $L$ of the interior region it
occupies. Moreover, such a length is quantized, $L=2N\sqrt{\theta}$, in terms
of a minimum length scale $\sqrt{\theta}$. The approach has the advantage that
the black hole density is not a function of the mass (as is the case in some
recent treatments). The density is, instead a fixed quantity. As such, the
approach puts an upper bound on black hole density, making it a universal
parameter. The picture that emerges is that a black hole defined on a
non-commutative background is both non-singular, holographic and quantized.
Yet we also find, interestingly, that when taken over $L$ the average value of
the associated stress-energy tensor of the fields satisfies all classical
energy conditions of GR.

\end{abstract}

\section{Introduction}

General Relativity (GR) rests on the proposition that spacetime and matter
have an intertwined relationship; one in which matter instructs spacetime how
to curve while spacetime tells matter how to move [1]. According to GR, under
rather extreme conditions on their density, matter fields should induce
curvature singularities on spacetime. Whether it is a (future-directed) black
hole type or a (past-directed) cosmological type, a curvature singularity
renders spacetime geodetically incomplete [2]. Thus with regard to the
dynamics of matter fields, the initial singularity in cosmology would have
constituted a "dead start" just as a black hole singularity would constitute a
"dead end". Matter just can't easily wiggle its way out of a singularity! This
concern has led to the view that under such extreme conditions, the breakdown
is one of GR theory rather than one of spacetime. Approaches to rid the black
hole spacetime of physical singularities by modifying GR have a long history
that goes back to Bardeen [3-11]. Lately, the issue (or non-issue) of the
spacetime singularity is actively discussed in modern formulations of quantum
gravity such as\ of String Theory [12] and Loop Quantum Gravity (LQG) [13]. In
LQG, for example, the Ashtekar Hamiltonian [14], when quantized, leads to a
minimum area $a_{0} $ [15] which, in turn, has led to a LQG inspired
description of a non-singular Schwarzschild black hole interior [16-19] and
black hole evaporation process [20]. The two theories are however still in
evolution. For example, LQG still seeks a systematic approach to calculate
scattering amplitudes and cross-sections needed for quantitative results while
String Theory still seeks to extend the black hole description beyond
near-extremal cases [21].

A potential ingredient of possibly any future theory of quantum gravity [22],
[23] is the proposition that at very high (possibly Planckian) levels of
resolution, the geometry of spacetime is inherently non-commutative (for a
review see [21] and citations therein). In non-commutative gravity the
coordinates $x^{\mu}$ become operators which obey some non-commutational
relationship of the form $\left[  x_{\mu},x_{\mu}\right]  =i\theta_{\mu\nu}$,
where $\theta_{\mu\nu}=i\epsilon_{\mu\nu}\theta$ is a non-symmetric tensor
with dimensions of length-squared. An uncertainty relationship of the form
$\left\vert \Delta x_{\mu}\Delta x_{\nu}\right\vert \geq\frac{1}{2}\left\vert
\theta_{\mu\nu}\right\vert $ is then implied and the coordinates have a fuzzy,
smeared expectation value.\ \ \ \ \ \ \ \ \ \ \ \ \ \ \ \ \ \ \ \ \ \ \ \ \ \ \ \ \ \ \ \ \ \ \ \ \ \ \ \ \ \ \ \ \ \ \ \ \ \ \ \ \ \ \ \ \ \ \ \ \ 

The concept of a quantized spacetime has been around for well over seventy
years now [24]. Lately, an interest in non-commutative spacetimes has been
rekindled by studies in String Theory [21]. In black hole physics, the idea
that non-commutativity renders point-like objects to be smeared out has been
explored to remove the would-be black hole singularity [25-32] and even to
discuss gravastars [31]. In the more recent treatments the approach has been
to apply a Voros map [34] on a Dirac delta function-like central density. This
approach "softens" the otherwise singular density to a Gaussian form
$\rho_{\theta}=\frac{M}{\left(  4\pi\theta\right)  ^{\frac{3}{2}}}\exp\left(
-\frac{r^{2}}{4\theta}\right)  $. The result is that the stress-energy tensor
$T^{\mu\nu}$, and hence the spacetime curvature, is rendered finite everywhere
in the black hole interior.\ In this respect the procedure just described has
been successful with regard to regularization of matter distribution inside
the black hole interior. There are however, in our view, some open questions:\ \ \ \ \ 

(1) In the above Gaussian form of the density function the maximum density
$\rho_{\max}$ scales with the mass $M$. Noting that the black hole mass
spectrum covers a large parameter space, with a conservative estimate of
$\sim42$ orders of magnitude (from say a $1g$ primordial black hole to a
$10^{42}g$), this suggests an equally large spectrum in the central black hole
density. Thus, in theory the density function could take on any values since
neither theory nor observation puts a hard upper bound on black hole mass,
yet. On the other hand, one would hope instead that a non-singular black hole
should be defined by a universally fixed density parameter.

(2) A second question regards a technique whose application is not just
restricted to the above-mentioned approach. It is often the case that in order
to create a non-singular spacetime, one introduces fields whose stress-energy
tensor $T^{\mu\nu}=diag\left[  \rho,p_{r},p_{\bot},p_{\bot}\right]  $
satisfies a radial equation of state of the form $p_{r}=-\alpha\rho$, where
$\frac{1}{3}<\alpha\leq1$. As is known such a construction will generally
produce the desired negative pressure to off-set singularity formation.
However, a couple of questions arise from here: (i) Does the entire fluid
inside the black hole satisfy such an equation of state, as is often implied?
And if so, is the \textit{apriori} assumed dominance of gravity, and the
implied outer Schwarzschild horizon, still justified? Or if not, should one
not make a clear statement regarding the state (dust, stiff...?) and the
available fractional quantity of the matter fields, and how the interface of
the implied 2-fluid is defined, if one exists? (ii) In the absence of a full
quantum theory of gravity to deal with the issue definitely, one often
introduces the above equation(s) of state by hand, and one often justifies the
choice by the end result. It would help if a reasonably sound, albeit
phenomenological, motivation for the choice of such equation(s) of state could
be advanced in which one attempts to go a little beyond the "result-based" approach.

In this work, we seek to construct a non-singular black hole model (on a
non-commutative background) that also addresses the above-mentioned questions,
while not compromising the universality of the minimum length parameter
$\sqrt{\theta}$. In our approach we introduce (with justification) a 2-fluid
stress-energy tensor whose trace (curvature scalar) takes on different signs
at different but specific points in the black hole interior region occupied by
matter-energy. In general, the pressure components are non-vanishing at each
point in the region occupied by matter fields. However as we show the average
value of the individual pressure components across the region vanish so that
only the density parameter contributes to the full mass-energy. As a result,
the bulk matter interior of the black hole is shown to obey all the classical
energy conditions, even though these conditions are violated in various
specific regions in the same interior. Further, we will construct a black hole
whose interior density is independent of the black hole mass but is instead a
fixed universal parameter for the entire black hole mass-spectrum.

The rest of the paper is arranged as follows. In Section 2 we introduce matter
fields on a non-commutative background. Section 3 addresses the two issues
identified above for discussion and also presents the solution for the entire
spacetime. Features of a black hole on a non-commutative background in our
treatment are pointed out in Section 4. We conclude with a discussion in
Section 5.

\section{Matter fields on non-commutative background}

A field $f\left(  x\right)  $, when defined on a non-commutative manifold
suffers a deformation. Historically, there are two different prescriptions for
defining the map onto such manifolds [34], namely the Wick-Voros product
$f\left(  x\right)  \bigstar_{V}g\left(  x\right)  $ and the Moyal product
$f\left(  x\right)  \bigstar_{M}g\left(  x\right)  $. On a complex plane
$z_{\pm}=\frac{x^{1}\pm ix^{2}}{\sqrt{2}}$ the two products can be
respectively defined by $f\left(  x\right)  \bigstar_{V}g\left(  x\right)
=\underset{n}{\sum}\left(  \frac{\theta^{n}}{n!}\right)  \partial_{+}%
^{n}f\partial_{-}^{n}g=fe^{\overset{\leftarrow}{\theta\partial_{+}}%
\overset{\rightarrow}{\partial_{-}}}g$ and $f\left(  x\right)  \bigstar
_{M}g\left(  x\right)  =fe^{\frac{1}{2}\theta\left(  \overset{\leftarrow
}{\partial_{+}}\overset{\rightarrow}{\partial_{-}}-\overset{\leftarrow
}{\partial_{-}}\overset{\rightarrow}{\partial+}\right)  }g$, where
$\partial_{\pm}=\frac{\partial}{\partial z_{\pm}}$. While the two maps do
appear different, it has been shown [34] that the resulting S matrix from both
prescriptions is essentially the same.

In this work we take the matter fields towards the end of gravitational
collapse to constitute a homogeneous fluid condensate. We then point out how
the effects of non-commutativity will halt collapse to avoid formation of a
spacetime singularity. The resulting density field $\rho$ will be shown to be
independent of the black hole mass $M$. As a consequence the approach may be
used to describe the interior of a black hole of any mass $M$ in a consistent
way and at the same time to establish the maximum density as a universal parameter.

The problem, here, reduces to one of projecting a homogeneous density field
$\rho_{0}$ onto a spherical non-commutative manifold in a region $\mathcal{R}
$ of size $L$ and inter-lattice interval $2\sqrt{\theta}$, corresponding to a
lattice cell of size $\sqrt{\theta}$. In this work we will restrict the
effects of non-commutativity in the radial direction. We begin by finding it
convenient to write the density as a scalar product of a complex field
$\rho_{0}=\phi\phi^{\ast}$, which when mapped onto a non-commutative manifolds
deforms to $\phi\left(  z_{+}\right)  \bigstar\phi^{\ast}\left(  z_{-}\right)
$. Independent of the map used (Moyal or Voros) it is easily seen that
inducing a constant, homogeneous density field onto a non-commutative
background reproduces the constant field,
\begin{equation}
\phi\left(  z_{+}\right)  \bigstar_{V}\phi^{\ast}\left(  z_{-}\right)
=\phi\left(  z_{+}\right)  \bigstar_{M}\phi^{\ast}\left(  z_{-}\right)
=\phi\phi^{\ast}=\rho_{0}. \tag{2.1}%
\end{equation}
At first, it would appear as if the mapping exercise above has no effect,
since indeed the density $\rho_{0}$ remains unchanged. However, such an
appearance would be deceptive. To see why, consider the following questions:
(i) What are all the expected effects of a map on a non-commutative
background? (ii) Is the resulting field still homogeneous, for example? In
general based on the operator character of the map (Moyal or Voros, see above)
a field (or fields) introduced onto a non-commutative manifold (1) can
(self)-interact and/or (2) interact with the manifold on which they are
induced. In our case the initial field is homogeneous and therefore will not
self-interact (as Eq. 2.1 shows). However, it will expectedly interact with
(\textit{as it seeks to conform to}) the non-commutative background on which
it is projected. This latter effect, (to be discussed below) is not manifest
in Eq. 2.1.\ \ 

\begin{lemma}
A non-commutative spacetime manifold contains lattice (on-cell) and
inter-lattice (off-cell) regions. On-cell/(off-cell fields) manifest
characteristics reminiscent of on-shell/(off-shell) fields, respectively (with
respect to their motion equations).
\end{lemma}

\begin{proposition}
\bigskip In order to conform with its background, a homogeneous field mapped
on a non-commutative spacetime manifold will undergo an internal deformation
resulting into two different inhomogeneous fields. The deformation mechanism
restricts regular matter fields to reside dominantly on-cell and non-matter
field to reside dominantly off-cell. The deformation mechanism leaves the net
density invariant.
\end{proposition}

We will be guided by the Lemma and Proposition statements above in modeling
features of a black hole interior. Note, however, the constraint demanded by
Eq.1 that at any point in $\mathcal{R}$ the net density must still be
point-wise $\rho_{0}$. We proceed by noting that one can write the complex
field $\phi$ in terms of the lattice spacing, on which it is projected, as
$\phi=\phi_{0}e^{ik\left\vert z\right\vert }$ and $\phi^{\ast}=\phi
_{0}e^{-ik\left\vert z\right\vert }$, where we identify $k=\frac{\pi}%
{2\sqrt{\theta}}$, and where $2\sqrt{\theta}$ is the inter-lattice interval
corresponding to a cell size $\sqrt{\theta}$. This can simply be rewritten as
$\phi\phi^{\ast}=\left(  \phi_{0}\right)  ^{2}\left[  \cos k\left\vert
z\right\vert +i\sin k\left\vert z\right\vert \right]  \left[  \cos k\left\vert
z\right\vert -i\sin k\left\vert z\right\vert \right]  $, so that\ \ \ \ \ \
\begin{equation}
\rho_{0}=\rho_{0}\sin^{2}k\xi+\rho_{0}\cos^{2}k\xi, \tag{2.2}%
\end{equation}
with $\xi=\left\vert z\right\vert $ and with $\rho_{0}$ the density field on
the non-commutative back ground given by Eq. 2.1. The relation in Eq. 2.2 is
certainly a trivial mathematical identity which however demands a non-trivial
physical interpretation.

We interpret the above result in the following manner.\ First, we consider the
non-commutative region $\mathcal{R}$ in which the fields reside to be a
spherical lattice of radial size $L=2N\sqrt{\theta}$ where $N=1,2,...$. By the
Proposition, the map $\phi\left(  z_{+}\right)  \bigstar\phi^{\ast}\left(
z_{-}\right)  $ can be viewed to constitute a mechanism through which the
original matter density field deforms over the region $\mathcal{R}$ into two
physically different fields, which we denote as $\rho_{m}$ and $\rho_{d}$. We
recognize the two terms on \ the right hand side of Eq. 2.2 to represent the
two fields and assign \textit{\ }%
\begin{align}
\rho_{m}  &  =\rho_{0}\sin^{2}k\xi,\ \nonumber\\
\rho_{d}  &  =\rho_{0}\cos^{2}k\xi. \tag{2.3}%
\end{align}
The field $\rho_{m}$ resides predominantly on-cell and is therefore, by the
Proposition, a regular matter density field. The field $\rho_{d}$ resides
predominantly off-cell. Note, however, that at any point in the black hole
region $\mathcal{R}$ the net density $\rho$, which satisfies $\rho_{0}%
=\rho_{m}+\rho_{d}$, is still manifestly constant after the map on the
non-commutative background (consistent with Eq. 2.1). We make the justifiably
common choice that the matter field is stiff, obeying an equation of state
$p_{m}\left(  \rho_{m}\right)  =\rho_{m}$. Based on this choice of matter
fields equation of state and based on the off-shell character of $\rho_{d}$
implied by the Proposition, and based on (later) considerations of energy
conservation, one is left the only alternative that the off-cell density field
$\rho_{d}$ need obey an equation of state of the form $p_{d}\left(  \rho
_{d}\right)  =-\rho_{d}$. This latter field is therefore de
Sitter-like\footnote{We call the field de Sitter-like because, as pointed out
later, it does not possess the standard symmetry requirements of the de Sitter
group.}. We emphasize that this result, that the off-cell field be de
Sitter-like, is not merely introduced by hand (as is usually the case), but is
instead a consequence of the non-commutative deformation mechanism that we
propose. We therefore have, on use of Eq. 2.2, that in region $\mathcal{R}$
the respective pressure terms are given as,\ \ \ \ \ \
\begin{align}
p_{m}\left(  \rho_{m}\right)   &  =\rho_{0}\sin^{2}k\xi\ ,\nonumber\\
p_{d}\left(  \rho_{d}\right)   &  =-\rho_{0}\cos^{2}k\xi. \tag{2.4}%
\end{align}
The results in Eq. 2.4 imply that at any point in region $\mathcal{R}$ the net
pressure in the $\xi$-direction, $p=p_{m}+p_{d}$, associated with an
inhomogeneous 2-fluid with a net density given by Eq. 2.2 satisfies an
equation of state,
\begin{equation}
p_{\xi}=-\rho_{0}\cos2k\xi. \tag{2.5}%
\end{equation}

\bigskip These results are represented in Figures 1 and 2, for an exemplary
black hole of size $L=6\sqrt{\theta}$ or $N=3$. Figure 1 displays the density
fields $\rho_{m}$ and $\rho_{d}$ and the net density $\rho=$ $\rho_{m}%
+\rho_{d}$. Figure 2 displays the corresponding pressures $p_{m}$, $p_{d}$ and
the net pressure $p=$ $p_{m}+p_{d}$. Note that while the net density is
generally constant (except at the outermost edge), the net pressure at any
point is a function of the coordinate $\xi$.\ \ \ %

%TCIMACRO{\FRAME{ftbpFU}{5.1309in}{2.1292in}{0pt}{\Qcb{The density components
%(check labels): $\rho_{m}$ of the matter fields, $\rho_{d}$ of the de
%Sitter-like field, and $\rho_{0}$ the net density in the interior region
%$\mathcal{R}$ of a size $N=3$ black hole.}}{}{l6c76a00.eps}%
%{\special{ language "Scientific Word";  type "GRAPHIC";  display "USEDEF";
%valid_file "F";  width 5.1309in;  height 2.1292in;  depth 0pt;
%original-width 6.8476in;  original-height 5.7449in;  cropleft "0";
%filename 'L6C76A00.eps';file-properties "XNPEU";}}}%
%BeginExpansion
\begin{figure}
[ptb]
\begin{center}
\includegraphics[
height=2.1292in,
width=5.1309in
]%
{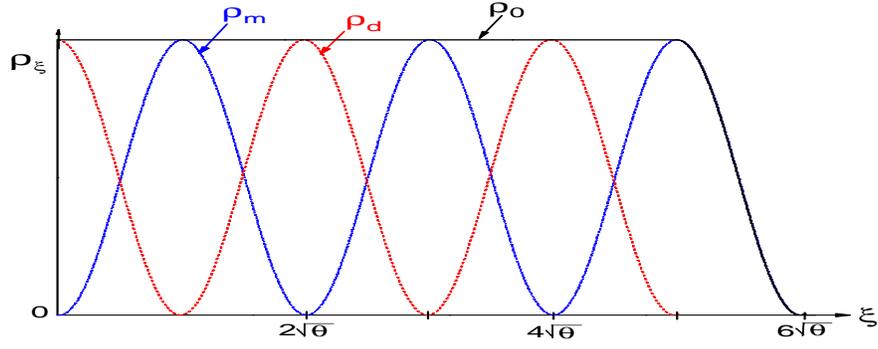}%
\caption{The density components (check labels): $\rho_{m}$ of the matter
fields, $\rho_{d}$ of the de Sitter-like field, and $\rho_{0}$ the net density
in the interior region $\mathcal{R}$ of a size $N=3$ black hole.}%
\end{center}
\end{figure}
%EndExpansion
In the tangential directions, the pressure $p_{\bot}$ of the isentropic fluid
satisfies an equation of state $p_{\bot}=p_{\xi}-\frac{1}{2}\xi\partial_{\xi
}p_{\xi}$ which in this case is given by $p_{\bot}=\rho_{0}\left[  \xi
k\sin2k\xi-\cos2k\xi\right]  $. We find it useful in our treatment to relax
this condition to $p_{\bot}=p_{\xi}-\frac{1}{2}\xi\partial_{\xi}\rho_{\xi}$.%

%TCIMACRO{\FRAME{ftbpFU}{5.3186in}{2.7639in}{0pt}{\Qcb{The pressure components
%(check labels): $p_{m}$ of the matter fields, $p_{d}$ of the de Sitter-like
%field, and $p$ the net pressure in the $\xi$ direction in the interior region
%$\mathcal{R}$ of a size $N=3$ black hole}}{}{l6c76a01.eps}%
%{\special{ language "Scientific Word";  type "GRAPHIC";  display "USEDEF";
%valid_file "F";  width 5.3186in;  height 2.7639in;  depth 0pt;
%original-width 6.8476in;  original-height 6.3711in;  cropleft "0";
%filename 'L6C76A01.eps';file-properties "XNPEU";}}}%
%BeginExpansion
\begin{figure}
[ptb]
\begin{center}
\includegraphics[
height=2.7639in,
width=5.3186in
]%
{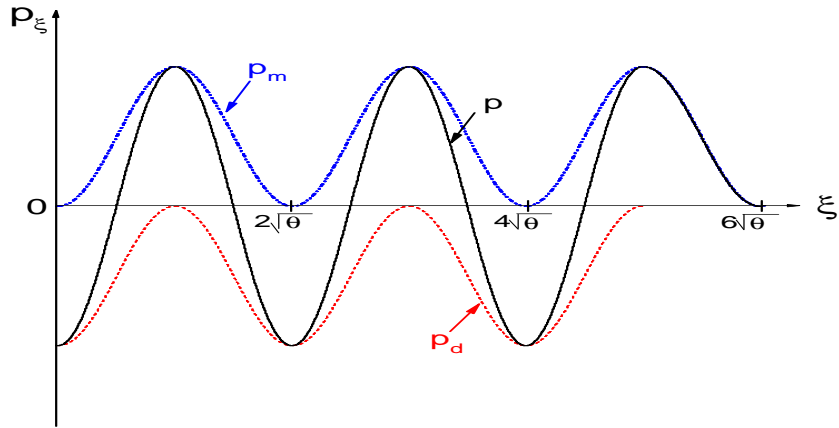}%
\caption{The pressure components (check labels): $p_{m}$ of the matter fields,
$p_{d}$ of the de Sitter-like field, and $p$ the net pressure in the $\xi$
direction in the interior region $\mathcal{R}$ of a size $N=3$ black hole}%
\end{center}
\end{figure}
%EndExpansion
In this form the condition leaves the stress-energy tensor isotropic, since
$\partial_{\xi}\rho_{\xi}=0$. Notice though that while the pressure is
isotropic, the 2-fluid is still inhomogeneous.\ The stress-energy tensor
$T^{\mu\nu}=diag\left[  \rho_{0},p_{\xi},p_{\bot},p_{\bot}\right]  $ for this
spacetime then takes the form
\begin{equation}
T^{\mu\nu}=diag\left[  \rho_{0},-\rho_{0}\cos2k\xi,-\rho_{0}\cos2k\xi
,-\rho_{0}\cos2k\xi\right]  . \tag{2.6}%
\end{equation}

\section{The Spacetime}

\subsection{Matter distribution, and the density parameter\ \ \ }

A major motivation of this work has been to discuss how matter may be
distributed inside a black hole. In particular, the two issues were to
construct a reasonably justified stress-energy tensor of the fields and also
to make the black hole density a fixed parameter by making it independent of
the mass. In this sub-section we briefly discuss these two issues, before
presenting the spacetime solution.\ \ 

The interior sub-region\ $\mathcal{R}$, where the fields reside, constitutes
bands of concentric spheres of thickness$\sqrt{\theta}$ alternatingly
dominated by matter and de Sitter-like fields, respectively. Each matter fluid
region always encloses a de Sitter-like fluid region. The alternating set-up
goes all the way down to the innermost central sphere $0\leq\xi\leq\frac{1}%
{2}\sqrt{\theta}$, which sphere (see Figure 1 and 2) is de Sitter-like. At any
2-surface $\xi$ in the region $0\leq\xi\leq\left(  L-\sqrt{\theta}\right)
\approx L$ the density of the black hole has a constant value $\rho_{0}$. Note
that for most black holes $L>>\sqrt{\theta} $ ( the smallest black hole,
$N=1,$\ $L=2\sqrt{\theta}$) so that for the purposes of discussing the density
profile, the edge effects can be neglected and virtually $\rho\left(
\xi\right)  =\rho_{0}$ for $0\leq\xi\leq L$. Further, we have shown that the
negative pressure (at various specific values of $\xi$) responsible for
halting the formation of a curvature singularity is produced through the
non-commutative deformation mechanism which we proposed. As is seen in Figure
2 the net pressure oscillates between positive and negative values due to the
on-cell matter fields and off-cell de Sitter-like field. Moreover, across the
region $\mathcal{R}$ occupied by the matter fields (see Figure 2) the average
value of the pressure elements $\left\langle p_{\xi}\right\rangle $ (and hence
also of $\left\langle p_{\bot}\right\rangle $) is a vanishing quantity when
taken over the entire $L$. Consequently, the only global contribution of the
stress-energy tensor to the total energy is that coming from the density
parameter $\rho_{0}$ and given by
\begin{equation}
\langle T_{\ \nu}^{\mu}\rangle=diag\left[  -\rho_{0},0,0,0\right]  \tag{3.1}%
\end{equation}
This result in Eq.3.1 implies that the average value of the stress-energy
tensor of this spacetime (and hence the bulk of matter in the interior black
hole region $\mathcal{R}$) obeys all the classical energy conditions of GR,
while Eq. 2.6 shows that specific surfaces in this region do not. In this way
the non-commutativity driven local quantum behavior gives way to global
classical behavior. The discussion here, along with our previous construction
of the energy-stress tensor (which was motivated by the proposed deformation
mechanism) addresses one of the issues we set out to deal with, namely the
issue of the choice of the stress-energy tensor and the associated equation(s)
of state.

That the\ only global contribution of the stress-energy tensor to the total
energy is that coming from the density parameter $\rho_{0}$ implies that the
deformation of the matter fields by the non-commutative background is, in this
sense, energy conserving and allows one to recover the original mass as
$M=4\pi\int_{0}^{L}\rho_{0}\xi^{2}d\xi$ or\ \ \
\begin{equation}
M=\frac{4}{3}\pi\rho_{0}N^{3}\theta^{\frac{3}{2}}. \tag{8}\label{3.2}%
\end{equation}
This mass $M$ is spread in the region $\mathcal{R}$ bounded by a 2-surface of
size $L^{2}=\left(  2N\sqrt{\theta}\right)  ^{2}$where $N=1,2...$ is the only
variable in the relation. Clearly the mass defines its size $L=L\left(
M\right)  $. In turn this implies that the density $\rho_{0}$ is independent
of the mass, and is in our case constant. We can thus treat $\rho_{0}$ as an
upper bound of any matter density due to gravitational collapse, and therefore
as a universal parameter. The result (Eq. 3.3) departs from results in which
the black hole of any mass is contained inside a fixed area $\theta$.
Consequently, in our work, there is no degeneracy in mass-confinement of a
black hole. This feature that a black hole determines its length size, with
the result that the black hole density is a universal parameter, is the last
of the two issues we set out to address.

Before leaving this discussion on matter distribution inside a black hole, it
is useful to point out how the two fields which constitute the matter-energy
behave at both boundaries $\xi=0$ and $\xi=L$, and how this behavior affects
the solution. As both Eqs. 2.2-2.5 and Figure 2 show, at the center $\xi=0$,
the matter fields vanish and the de Sitter-like field maximizes, to pure de
Sitter, with a fixed density $\rho_{0}$ and pressure $-\rho_{0}$. It is the
negative pressure in this innermost region $0\leq\xi\leq\sqrt{\theta}$ that
halts the immediate outlying spherical matter region $\sqrt{\theta}\leq\xi
\leq2\sqrt{\theta}$ from forming a curvature singularity at the center. On the
other hand, at the boundary $\xi=L$ of $\mathcal{R}$ the fields vanish in a
well-behaved manner (in the mathematical sense) and for $\xi\geq L$ no such
fields exist. This feature implies smooth continuity of the metric across the
matter vacuum boundary, as demanded by the Israel junction conditions [36] [35].

\subsection{The spacetime metric}

The general region of a black hole inside the horizon is a Kantowski-Sachs
spacetime, in which the radial and temporal coordinates switch their roles so
that $r\rightarrow t$ and $t\rightarrow r$. As a result the variable
$\xi=\left\vert z\right\vert $ in the preceding equation can now be identified
(for $0\leq\xi<2M$) as a temporal variable $\xi=t$, where $t=0$ is measured
from the "center" of the spherical geometry of the spacetime.\ \ \ \ 

In the sub-region $\mathcal{R}$ the fields at any 2-sufarce $\xi$ are
described by the stress-energy tensor $T_{\ \nu}^{\mu}=diag\left[  -\rho
_{0},p_{\xi},p_{\bot},p_{\bot}\right]  $ given by Eq. 2.6. On solving the
Einstein Field Equations [34] subject to Eq. 2.6 one finds the region
$\mathcal{R}$ to admit a metric line element,\ \ \ \ \
\begin{equation}
ds^{2}=-\left(  1-\frac{2m\left(  \xi\right)  }{\xi}\right)  ^{-1}d\xi
^{2}+\left(  1-\frac{2m\left(  \xi\right)  }{\xi}\right)  d\xi^{2}+\xi
^{2}d\Omega^{2},\tag{3.3}%
\end{equation}
where $m\left(  \xi\right)  $ is the mass enclosed by the coordinate $\xi$ in
the region $0$ $\leq\xi\leq L$. At the matter-vacuum inter-surface $\xi=L$,
$m\left(  \xi\right)  \rightarrow M$.

The outer interior sub-region $L\leq\xi\leq2M$ which is still Kantowski-Sachs
satisfies the vacuum Einstein equations to give a line element\ \
\begin{equation}
ds^{2}=-\left(  1-\frac{2M}{\xi}\right)  ^{-1}d\xi^{2}+\left(  1-\frac{2M}%
{\xi}\right)  dr^{2}+\xi^{2}d\Omega^{2}.\tag{3.4}%
\end{equation}
One notes that the functions representing the fields and their derivatives
vanish at the 2-surface $\xi=L$, and are therefore well-behaved thereat. Such
a feature guarantees the smooth matching of the metrics given by Eqs. 3.3 and
3.4 as they satisfy the Israel junction conditions [35], [34].

Finally at the outer, or Schwarzschild horizon $\xi=2M$, the spatial and
temporal coordinates switch and for $2M<r<\infty$ we recover the exterior
vacuum Schwarzschild solution,%
\begin{equation}
ds^{2}=-\left(  1-\frac{2M}{r}\right)  dt^{2}+\left(  1-\frac{2M}{r}\right)
^{-1}dr^{2}+r^{2}d\Omega^{2}. \tag{3.5}%
\end{equation}

\section{ Features}

In the previous section we have discussed two features of the black hole in
our model, namely the local (quantum) and global (classical) behavior of the
fields in the energy-stress tensor and the mass independence of the density
parameter. In this section we briefly point out more features of the black
hole in this model that we consider essential.

\subsection{A black hole is a quantized, non-singular object}

We have seen that the matter inside a black hole is distributed in a region of
size $L=2N\sqrt{\theta}$, $N=1,2,...$. This length size $L$ is manifestly
quantized in $\sqrt{\theta}$, as a result of non-commutative considerations.
Thus, in this model, a black hole interior constitutes bands of quantized
concentric spheres, each of average thickness$\sqrt{\theta}$, alternately
dominated by matter and de Sitter-like fields, respectively. Each matter fluid
region always encloses a de Sitter-like fluid region. The alternating set-up
goes all the way down to the innermost central sphere. This last innermost
sphere is (see Figure 1 and 2) manifestly de Sitter-like. The tendency for a
given spherical band of matter fluid to collapse under its gravitational
influence is therefore always counter-balanced by the negative pressure of the
bounding inner sphere of de Sitter-like fluid. The result is that formation of
the central singularity due to gravitational collapse is avoided. Based on
this set-up, nature appears to protect spacetime from formation of curvature
singularities. This feature is consistent with a conjecture that we previously
put forward [37].\ 

\subsection{Trapped surfaces and horizons}

The region $\mathcal{R}$ consists of a system of nested trapped and untrapped
surfaces. For example $\xi=L$, the outermost matter surface closest to the
Schwarzschild horizon, is a trapped surface. On the other hand the central
point $\xi=0$ which lies on a de Sitter space (see Eqs. 2.2-2.5) is untrapped.
Based on this and considering, as a simple example, the spacetime of the
smallest black hole ($L=2\sqrt{\theta}$, $N=1$) of some mass $M$. Such a
spacetime will have one region that constitutes a set of trapped surfaces
bounded from out by $\xi<2M$, and one inner region of untrapped surfaces that
includes the black hole center. In between, the matter density pick and the
adjacent de Sitter-like density pick at the center there exits some marginally
trapped surface centered at $\xi=0$ that divides the trapped and untrapped
regions. Such a surface is a horizon. Thus the smallest black hole will have
two horizons, including the Schwarzschild horizon. For now we will defer a
detailed analysis of the horizon structure of this spacetime for a future discussion.

\subsection{A black hole is holographic}

The picture of a black hole interior that has emerged so far is one of a
system of concentric spherical bands of matter fields interjected by
concentric spherical bands of de Sitter-like regions. This quantized
onion-like system goes in all the way down to the center, with the central
region being de Sitter-like (see Eqs. 2.2-2.5 and Figures 1-3). Figures 1 and
2 show that the surface $\xi=\left(  N-\frac{1}{2}\right)  2\sqrt{\theta} $ is
special. We will call this 2-surface $\Sigma_{s}$. Surface $\Sigma_{s} $ is
the boundary inside which both (i) the black hole density is constant and
maximum $\rho=\rho_{0}$, and (ii) the entire de Sitter-like fluid is confined.
We consider the 2-fluid of matter and de-sitter-like field inside $\Sigma_{s}$
to constitute a frozen system, in the following sense. We suppose that during
gravitational collapse the fields crossing $\Sigma_{s}$ will deposit all their
dynamical degrees of freedom in the thin sub-region $\left(  N-\frac{1}%
{2}\right)  2\sqrt{\theta}\leq\xi\leq N2\sqrt{\theta}$ overlying $\Sigma_{s}$,
which we will call sub-region $\Delta$. Consequently, the 2-fluid inside
$\Sigma_{s}$ is a super-frozen condensate with non-existent dynamical degrees
of freedom and hence with zero entropy.

One recognizes sub-region $\Delta$ overlying the condensate system to be the
only region (i) made purely of matter fields (ii) with a net density $\rho$
that monotonically decreases to zero and (iii) with a free boundary at
$L=N2\sqrt{\theta}$. The free outer boundary character\ of sub-region $\Delta$
facilitates the existence of quantum fluctuations which we suppose dominate
this region. Thus, based on this argument, all dynamical degrees of freedom of
the matter fields in a black hole will reside in sub-region $\Delta$. It is
important to recognize that in the $\xi$ direction the sub-region $\Delta$
where quantum fluctuations are likely to dominate has a thickness, or more
precisely a skin-depth, of size $\sqrt{\theta}$. Recall that the smallest
length measure allowed by the uncertainty relation suggested by
non-commutative gravity is $\left[  \sqrt{\left\vert \Delta x_{\mu}\Delta
x_{\nu}\right\vert }\geq\sqrt{\frac{1}{2}\left\vert \theta_{\mu\nu}\right\vert
}\right]  $. Based on this, it is clear that sub-region $\Delta$ cannot be
distinguished from the surface $L=2\sqrt{\theta}$. As a result we find that in
our treatment the dynamical degrees of freedom of a black hole can only exist
on the surface $\xi=L$ and will therefore scale with the area of the surface
as $L^{2}\sim4N^{2}\theta$. Moreover, such information restricted on outermost
matter surface $L$ from below can be propagated\ onto the Schwarzschild
horizon. Quantum fluctuations of the matter fields on the matter surface $L$
can set up sympathetic vacuum fluctuations in the overlying region $L\leq
\xi\leq2M$, which deposit information on the Schwarzschild horizon. In this
sense our treatment suggests a physical picture of how it can be that the
dynamical degrees of freedom (and hence the entropy of a black hole) reside on
the surface, consistent with the Holographic Principle. Based on this argument
it follows that, in this treatment, a black hole is holographic [38].

\section{Closing discussion}

In this work we have put forward a model of a black hole, set on a
non-commutative background. The entire spacetime is an exact solution of the
Einstein Field Equations. Our approach is based on a set of axiomatic
statements we make, namely (i) a proposition that imposes a non-commutative
deformation mechanism on a density field introduced on a non-commutative
background into two different fields, and\ (ii) a lemma that points to where
such fields will reside on the manifold. It is found that a de Sitter-like
field is one of the by-products of the deformation mechanism. Such a de
Sitter-like field which resides in the inter-lattice (off-cell) regions is
responsible for producing negative pressure that offsets singularity
formation. The other product, the matter field, resides on-cell. The result is
that the interior region\ $\mathcal{R}$, where the fields reside, constitutes
bands of concentric spheres of thickness $\sqrt{\theta}$ alternatingly
dominated by matter and de Sitter-like fields, respectively. Each matter fluid
region always encloses a de Sitter-like fluid region. The alternating set-up
goes all the way down to the innermost central sphere, which sphere is
manifestly de Sitter-like. This arrangement results in a non-singular,
quantized black hole. The approach has the advantage that the black hole
density is not a function of the mass (as is the case in some recent
treatments). The net density $\rho_{0}$ is, instead a fixed quantity. The
approach puts an upper bound on black hole density as a universal
parameter.\ This is one of the questions this work set up to address. \ \ \ \ \ \ \ 

The second question, regarding the stress-energy tensor has also been
addressed. It is shown that the negative pressure (at various specific values
of $\xi$) responsible for halting the formation of a curvature singularity is
produced through the non-commutative deformation mechanism which we proposed.
The choice of the equation of state and hence the choice of the stress-energy
tensor is justified only by the the proposed mechanism. We found, however,
that across the region $\mathcal{R}$ occupied by the matter fields the average
value of each of the pressure elements in the stress-energy tensor is a
vanishing quantity.\ The mechanism conserves the original energy in the mass
$M$. Interestingly, this result also implies that the average value of the
stress-energy tensor of this spacetime (and hence the bulk of matter in the
interior black hole region $\mathcal{R}$) satisfies all the classical energy
conditions of GR, while individual specific surfaces in this region do
not.\ The mixed result is important to the extent that it demonstrates locally
quantum characteristics of the fields due to a non-commutative background at
each surface $\xi$ sum up to present a globally classical view.

The spacetime proposed has an interior central region $\mathcal{R}$ of size
$L$ where the fields reside. The region $\mathcal{R}$ is bounded from above by
an interior vacuum region $L<\xi\leq2M$. The full sub-space $0\leq\xi\leq2M$
is a Kantowski-Sachs spacetime. On the other hand, the region $2M<\xi
\leq\infty$ is still the asymptotically flat Schwarzschild spacetime. The
entire spacetime is described by a well-behaved metric (continuous with
continuous derivatives) satisfying the Israel junction conditions. In
particular, the matter-vacuum interface $\xi=L$, is manifestly well-behaved.\ \ 

We have found that a black hole described on a non-commutative background
presents several new features and suggests physically motivated intuitive
interpretations to some others already known. In particular, we showed that a
black hole defines its own size scale $L=L\left(  M\right)  $ of the interior
region it occupies $\mathcal{R}$. This region is quantized in terms of
$\sqrt{\theta}$ as $L=2N\sqrt{\theta}$, where $N=1,2,3..$ and $\sqrt{\theta}$
is the non-commutative length scale (corresponding to the cell size), which in
this treatment turns out to be the size scale of each spherical band. Finally,
we pointed out that the matter interior of the black hole in this work is
frozen, with no existing dynamical degrees of freedom. Such degrees of freedom
are shown, with justification, to exist on the surface instead. In our view
this creates an intuitive basis for why the entropy scales with surface area
of the horizon.\ \ \ 

In summary we find that a black hole defined on a non-commutative background
is necessarily both non-singular, quantized and holographic. A discussion of
the horizon structure, radiation process and stability of this spacetime will
be reported elsewhere.\

\end{document}